\documentclass{llncs}  

\usepackage{graphicx} 
\usepackage[T1]{fontenc}
\usepackage{amsmath}
\usepackage[english]{babel}
\usepackage{array,tabularx}
\usepackage{booktabs} 
\usepackage{makeidx}  % allows for index generation
\usepackage{xparse}
\usepackage{moredefs}
\usepackage{lips}
\usepackage{epstopdf} 
\usepackage{color}
\usepackage{csquotes}
\usepackage{xcolor}
\usepackage{times}
\usepackage[hidelinks]{hyperref}
\usepackage[nolist,nohyperlinks]{acronym}
\usepackage{comment}
\usepackage{paralist}
\usepackage{cleveref}
\usepackage{wrapfig}
\usepackage{multirow}
\usepackage{microtype}
\usepackage{listings}
\usepackage[super]{nth}
\usepackage{natbib}

\usepackage{float}
\floatstyle{plaintop}
\restylefloat{table}

\makeatletter
\renewcommand\@biblabel[1]{#1.}
\makeatother
\addto\captionsenglish{}

\usepackage{fancyhdr}
\fancypagestyle{WI_footer}{\fancyhf{}\fancyfoot[L]{\color{black}\scriptsize submitted to 19th International Conference on Wirtschaftsinformatik\\September 2024, Würzburg, Germany}}

\crefformat{footnote}{#2\footnotemark[#1]#3}

\DeclareMathOperator{\compose}{{{\normalfont\text{\textbullet}}}}

\usepackage{xspace}

\expandafter\def\expandafter\UrlBreaks\expandafter{\UrlBreaks%  save the current one
  \do\a\do\b\do\c\do\d\do\e\do\f\do\g\do\h\do\i\do\j%
  \do\k\do\l\do\m\do\n\do\o\do\p\do\q\do\r\do\s\do\t%
  \do\u\do\v\do\w\do\x\do\y\do\z\do\A\do\B\do\C\do\D%
  \do\E\do\F\do\G\do\H\do\I\do\J\do\K\do\L\do\M\do\N%
  \do\O\do\P\do\Q\do\R\do\S\do\T\do\U\do\V\do\W\do\X%
  \do\Y\do\Z}

\newcolumntype{L}[1]{>{\raggedright\arraybackslash}p{#1}}   % linksbündig mit Breitenangabe
\newcolumntype{C}[1]{>{\centering\arraybackslash}p{#1}}     % zentriert mit Breitenangabe
\newcolumntype{R}[1]{>{\raggedleft\arraybackslash}p{#1}}    % rechtsbündig mit Breitenangabe

\begin{document}
\frontmatter          % for the preliminaries

\mainmatter              % start of the contributions

\title{Towards a Theoretical Foundation of Process Science}

%\author{Blinded\inst{1}}
%\authorrunning{Blinded et al.} % abbreviated author list (for running head)
%\institute{Blinded}
\subtitle{Research in Progress} %Specify type of research paper here!
\author{Peter Fettke\inst{1,2} \and
Wolfgang Reisig\inst{3}}

\institute{German Research Center for Artificial Intelligence (DFKI), Saarbrücken, Germany\\
\email{peter.fettke@dfki.de} \and
Saarland University, Saarbrücken, Germany \and
Humboldt-Universität zu Berlin, Berlin, Germany\\
\email{reisig@informatik.hu-berlin.de}}

% -----------------------
% |  Begin of Document  |
% -----------------------
\maketitle
\setcounter{footnote}{0}

% ------------- 
% |  Abstract and Keywords  | 70 bis 150 words
% -------------
\begin{abstract}
Process science is a highly interdisciplinary field of research. Despite numerous proposals, process science lacks an adequate understanding of the core concepts
of the field, including notions such as process, event, and system. A more systematic framework to cope with process science is mandatory. We suggest such a framework using an example. The framework itself addresses three aspects: architecture, statics, and dynamics. Corresponding formal concepts, based on established scientific theories, together provide an integrated framework for understanding processes in the world. We argue that our foundations have positive implications not only for theoretical research, but also for empirical research, e.g., because hypothesized relationships can be explicitly tested. It is now time to start a discussion about the foundations of our field.\\\\

\end{abstract}

\thispagestyle{WI_footer}

% ------------- 
% |  Content  |
% -------------

\section{Introduction}\label{sec:Introducion}

Like numerous other terms in computing disciplines, the term “business process management” (BPM) is used to denote (a) a particular kind of \emph{problem}, (b) a possible \emph{solution} to that problem, and (c) the \emph{academic field} that studies the problem and its known solutions:

%\begin{itemize}

(a) BPM as a \emph{problem}: People, organizations, and systems run processes in many domains, such as manufacturing, healthcare, or logistics. It is an undeniable fact that in these domains resources are limited. Therefore, managing processes is essential in almost all domains. In other words, the analysis, planning, design, implementation, and monitoring of processes are practical problems in the daily work of many people in almost every organization.

(b) BPM as a \emph{solution}: In both academia and industry, numerous ideas have emerged to deal with BPM. The proposed solutions include: (a) general frameworks, e.g., the six core framework of \cite{rosemann2010sixcore}, (b) lifecycle models, e.g., \cite{dumas2018fundamentals}, (c) modeling approaches, e.g., BPMN, (d) specific algorithms, e.g., algorithms for process discovery, or (e) available software tools. Each of these artifacts can be interpreted as a solution to a BPM problem.

(c) BPM as an \emph{academic field} of research: BPM has a long tradition of research (\cite{beverungen2021bpm,houy2010:bpm_large,recker2016field}). Considering the wide and heterogeneous spectrum of various BPM problems and solutions, it is not surprising that the field is interdisciplinary by nature and is rooted in different academic disciplines. Some important branches of the field and their main outlets for publishing results are: \emph{Business Process Management Journal} (since 1995), \emph{International Conference on Business Process Management} (since 2003), or -- more specialized -- \emph{International Conference on Application and Theory of Petri Nets} (since 1980), \emph{International Conference on Process Mining} (since 2019). Furthermore, many major journals or conferences have special BPM sections, e.g. \emph{Business and Information Systems Engineering}, \emph{International} and \emph{European Conferences on Information Systems}, or \emph{International Conference on Wirtschaftsinformatik}. Last, but not least, the term “\emph{process science}” has recently been used as a name for the field (\cite{brocke2021science}) and for a newly founded journal.
%\end{itemize}

The problems, solutions, and branches of BPM provide a rich picture of different academic ideas that are important for our discipline. Against this background, the search for the theoretical foundations of process science is of great importance: What are the field's core concepts? What are the core phenomena? What are the assumptions of the scientific community? Which hypotheses are controversial?

This research explores the theoretical foundations of the field of process science. It proceeds as follows: After this introduction, Section 2 examines the fundamental problem in more detail. A well-accepted scientific background of our study is summarized in Section 3. Section 4, the main part of the paper, proposes a framework as a theoretical foundation of our field by means of an example. The following Sections 5 and 6 discuss related work and conclude our argumentation.

\section{The quest for theoretical foundations}

Process science lies at the intersection of technology-oriented and human-oriented disciplines, e.g., (management) information systems, (business) informatics, social sciences, and mathematics. All these disciplines contribute to a deeper understanding of the core concepts of the field (\cite{brocke2021science}).

On an \emph{intuitive level}, it is obvious what is meant by the notion of a process. However, making this intuitive notion explicit is far from being trivial. For example, a \emph{process} is often understood as a \emph{sequence} of events occurring in time. But when is a sequence of events a \emph{continuous} process? Moreover, the notion of sequence implies an overall, \emph{total} order. However, many phenomena in the real- or imagined-world can much better be understood as cause-effect relationships that are only \emph{partially} ordered.

Other theoretical questions arise, such as how \emph{objects} and \emph{data} are introduced into a theoretical framework. How do symbols of a language get their meaning? Is there a general notion of a \emph{schema} to describe the objects which are involved in a process? Is there a notion of a \emph{state} that can be used to characterize a process, especially its beginning or ending? Is such a notion based on a \emph{global} or \emph{local} understanding?

Finally, a process does not exist in isolation. It is clear that processes can interact with other processes. How can processes be \emph{composed}? Can the results be understood as a \emph{system} of processes? Or is there a diametrical difference between the understanding of a process and a system in the world?

In summary, the notion of a process and many related notions are intuitively clear. However, from a theoretical point of view, it is necessary to have a deeper understanding of the most basic concepts of process science. Today, it is far from clear how such an understanding can be achieved.

\section{The axiomatic method in empirical sciences as a framework}

The idea of axiomatizing a body of knowledge is anything but new and can be traced back, at least, more than 2,000 years to \emph{Euclid} who gives axiomatic foundations for geometry (\cite{Suppes_02}). Since then, tremendous progress has been achieved, e.g. the flourishing field of geometry develops in different branches with many important applications and different foundations, e.g. (non-) Euclidian geometry.
Besides particular applications in geometry, the idea of axiomatizing develops in different directions and is differently understood today. One widely accepted and well-practiced approach can be understood as an \emph{informal axiomatizing}. This approach is attributed as \emph{informal} because it does not use a formal system to formulate the main ideas of a body of knowledge. The framework used is based on \emph{intuitive set-theory} and combines formal with informal ideas (\cite{Suppes_57}). As such, it is always claimed that the formulated ideas can be expressed in the formal system of \emph{first order predicate logic}. However, in everyday scientific practice, just informal language is used.

The idea for this kind of axiomatizing can be linked to \emph{model theory}, as developed in logic and mathematics. Particularly, the \emph{Bourbaki} group applies this idea to overcome the basic crisis in mathematics at the beginning of the last century (\cite{Bourbaki_50}). However, in the meantime, this approach is the standard approach not only used in mathematics, but also in empirical sciences which numerous examples clearly and impressively demonstrate. For instance, \cite{Suppes_02} provides examples from physics, philosophy, psychology, computer science, economics, and semiotics.

The core idea of an informal axiomatizing is to introduce a \emph{structure} $S$ consisting of:

\begin{itemize}
\item \emph{Basic sets}: The basic sets define the basic objects of the structure $S$, e.g the symbol $M$ denotes a basic set.

\item \emph{Derived sets}: Further sets can be defined on basic sets or derived sets, e.g. the symbol $<$ denotes a binary relation on $M$.

\item \emph{Functions}: Mathematical functions can be defined as operations on sets, e.g. the symbol $f$ denotes a function $M$ on itself.

\item \emph{Properties}: Certain properties hold for all elements of the structure $S$ (“axioms”), e.g. for all $a$, $b$ element of $M$ holds: $a < b$ or $b < a$.
\end{itemize}

The axiomatic method is already intensively used in process science. Most, if not all \emph{computer science}-oriented branches of process science apply the axiomatic method. But, in \emph{management}-oriented approaches, this idea is already used, too, which is not only demonstrated by the quote of the Nobel Laureate Paul Milgrom:

{\footnotesize
“Formalization is important to economics, because it allows readers and others to identify the precise assumptions that underpin any purported conclusion, to verify that the assumptions really do imply this conclusion, and to check how deviations from the assumptions might alter the conclusion.” \cite[p.~3f.]{Milgrom2017}\par}

\section{An example to illustrate the proposed theoretical foundations}
Next, we highlight our proposal for the foundation of process science. We use a multi-branch restaurant as a \emph{case example} to sketch some of the core ideas of our proposal.

The interesting question is how the evolving dynamics of the world, the ongoing change, can be captured theoretically. To answer this question, we define a \emph{run} as a \emph{partially} ordered set of steps. Each step contains the occurrence of an \emph{event} along with the affected states. The most fundamental notions for understanding a run are the updating of states and the occurrence of events. These two notions are closely related: when an event occurs, some actually reached states are abandoned and some actually not reached states are reached.

\begin{figure}[tp]
    \centering
    \includegraphics[width=1.1\textwidth]{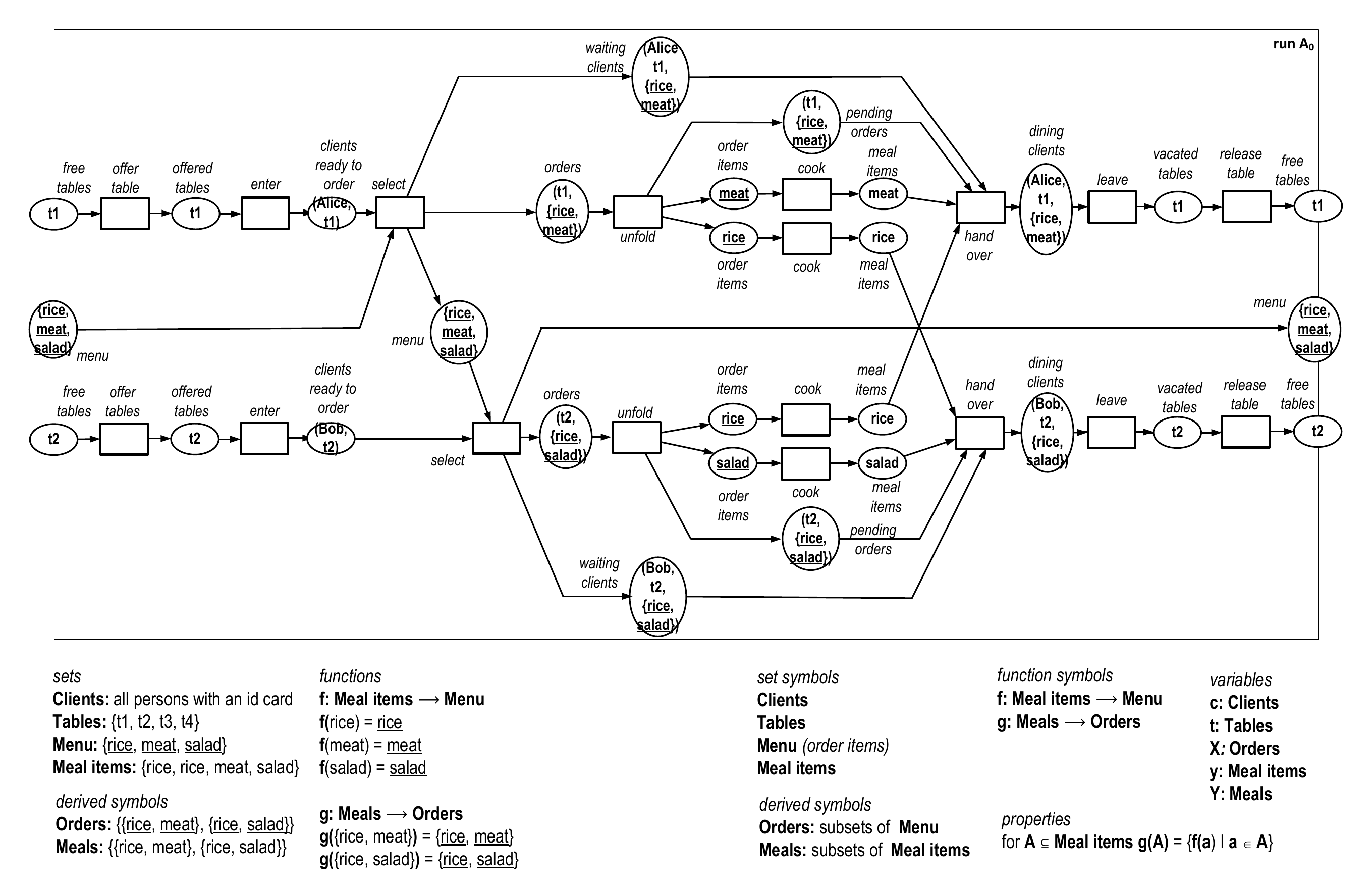}
    \caption{Core concepts of the example: run, structure, and signature}
    \label{fig:example}
\end{figure}

This intuitive idea can be formulated in terms of steps in a run. The top of Figure~\ref{fig:example} shows a particular flow of events: Alice and Bob independently enter the restaurant, order a meal, their meals are concurrently prepared and cooked, and served to both customers. After eating, the customers leave the restaurant. Note that some steps of the run are ordered, others are not, e.g. a dish is selected after a table is offered, but the two customers are offered tables and enter the dining area independently of each other. In other words, a run is not understood as a sequence of steps but as a \emph{partial} order of steps.

%\begin{figure}[htp]
%\begin{figure}[htp]
%    \centering
%    \includegraphics[width=1.2\textwidth]{figures/Ablauf_gesamt.png}
%    \includegraphics[width=.47\textwidth]{figures/Struktur.png}
%    \includegraphics[width=.49\textwidth]{figures/Signatur.png}
%    \includegraphics[width=.3\textwidth]{figures/drei_Module.png}
%    \caption{Core concepts of the example: run, structure, and signature}
%    \label{fig:example2}
%\end{figure}

Process science have to deal with \emph{real-} and \emph{imaginated-world items}, e.g. customers, products, as well as \emph{data}, e.g. information \emph{about} customers and products. Both aspects have to be integrated. To this end, we propose the standard notion of \emph{mathematical structure}. The most important static aspects of our example are shown in lower left corner of Figure~\ref{fig:example}, e.g. in this example some \emph{basic sets} are introduced to describe customers, tables, meals, and dishes. \emph{Derived sets} describe ordered dishes and cooked meals. Two functions specify which meals are cooked for certain ordered dishes.

So far we have described one particular run of a restaurant. Several runs represent the behavior in a restaurant. However, from a theoretical point of view, it is necessary to abstract from a particular restaurant. A more general understanding is needed. Our approach uses the classical ideas from predicate logic. Similarly, we distinguish between a \textit{system model} and a \textit{schema} for system models. Therefore, a \emph{signature} for all branches of the restaurant is introduced at the bottom right of Figure~\ref{fig:example}. The signature does not talk about particular sets, but about \emph{symbols} representing sets. These symbols can be used to describe events of a run on a schematic level. All branches of the restaurant in our example are structured according to the same schema and behave according to the same patterns; however, they differ in several other aspects. The schema of all branches, an example of a branch, and a run in that branch are different abstraction layers.
% The schema of all branches, an example of a branch, and a run in that branch are modeled; see \cite{fettke2022breathing} for details.

In addition, our approach uses a \emph{composition calculus} to provide a notion of the composition of modules. In our example, a branch is composed of three modules: \emph{branch} $=_{\text{def}}$ \emph{entry} $\compose$ \emph{dining area} $\compose$ \emph{kitchen}.

In summary, our proposal has three pillars:

\begin{itemize}

\item \emph{Architecture}: Today's computer-integrated systems are composed of multiple sub-systems. A module is the basic building block for composing complex systems. \emph{Composition} matters!

\item \emph{Dynamics}: Changes in the world are described by a partially ordered flow of events. \emph{Causality} matters!

\item \emph{Statics}: Structures describe symbolic objects as well as objects in the real- or imagined-world. \emph{Objects} matter!
\end{itemize}

Note, that all of the concepts in our proposal and in this example are introduced \emph{axiomatically}. More technically, our approach integrates well-known concepts, namely, the idea of \emph{model theory} and \emph{axiomatic specification} (statics), \emph{Petri nets} (dynamics), and \emph{composition calculus} (architecture). All three strongly integrated pillars provide a theoretical foundation for process science.

\section{Related work}

The classical process framework distinguishes between the notions of a \emph{process} $P$, sometimes also called \emph{system}, and a \emph{model} $M$, both are based on a set of \emph{activities} or \emph{events}, e.g. \cite{aalst2012process_mining}. A \emph{trace}, sometimes also called \emph{process instance}, is defined as a finite sequence of activities; a model $M$ and a process $P$ are both subsets of the set of all possible traces. In other words, classical \emph{automata theory} is used: $P$ and $M$ are understood as \emph{formal languages} over an \emph{alphabet}.

This foundation is typically used to understand processes. Such a theoretical conceptualization has an important advantage. However, it is well known that business processes contain further perspectives, e.g. \emph{objects}, \emph{data}, \emph{networked actions}, \emph{(sub-) systems}. Although there are already several alternative theoretical extensions of the classical framework, e.g. \cite{aalst2020object_centric, Fahland2019, tour2021agent}, it is an open question how these ideas can be used as a theoretical foundation for process science.

\cite{houy2011foundations,houy2014understandability} review the theoretical foundations for empirical research in BPM. The framework introduced by \cite{wand1990}, based on the philosophical work of Bunge, is one of the most prominent theoretical foundations from the information systems discipline. At the core, this approach follows the classical foundations of automata theory mentioned before.

In addition, there are many theoretical approaches which only offer ideas on a more or less \emph{intuitive level}, e.g. \cite{Scheer_12,Frank_14,Winter_01}. Although these works offer a rich picture of theoretical ideas, it is open how they provide \emph{explicit} theoretical foundations.

Another discussion focuses on what \emph{kind} of theoretical foundations are important. \cite{gregor2006nature} introduces different types of theory; \cite{bichler2016} provide some general discussion. It is not always clear whether these ideas are based on the well-accepted axiomatic method in empirical science on which our proposal is based on.

\section{Conclusions}

This paper emphasizes the need for theoretical foundations of process science. We strongly agree with the positive aspects of formalization mentioned in the quote from Nobel Laureate Paul Milgrom (cp. Section 3). Furthermore, we fully share the surprise and wonder described by \cite{weber1997} at the use formalization in our field:

{\footnotesize
“[W]e [Wand and Weber] have often been criticized for the formal approach we have used to articulate our models. I am perplexed by these criticisms because I cannot conceive of a discipline that is serious about its foundations if it proscribes use of mathematics to articulate these foundations.” \cite[p. 33]{weber1997}\par}

Looking back over the last decades, it is safe to say that many of the foundations introduced by Wand and Weber are well accepted and often empirically tested, especially in the information systems branch of process science. However, process science needs a broader discussion of foundations. Therefore, we believe that it is now time to intensify the discussion on the theoretical foundations of process science as a field of study.

This is especially important in the context of the digital revolution of the twenty-first century. In particular, it is necessary to clarify what is meant by digital processes and process technology, e.g. process mining and robotic process automation. Since several branches of process science have already started the discussion on how our field can embrace digital transformation, a deeper discussion is now needed.

In addition to the great advantages of the basic framework presented, we do not advocate abandoning other forms of scientific inquiry. The axiomatic method can easily be integrated with other methods such as bold speculation and empirical approaches as well as experiments and case studies. In this sense, we strongly believe in a bright future for our field.

%\section{Acknowledgements}
%You can add your acknowledgements to the revised paper. Acknowledgements are added before the references.

% ----------------
% | Bibliography |
% ----------------

\bibliographystyle{agsm}
\bibliography{main}
% ---------------------
% |  End of Document  |
% ---------------------

\end{document}